# Co-evolution of platform architecture, platform services, and platform governance: Expanding the platform value of industrial digital platforms

Marin Jovanovic [a],[*], David Sjödin [b],[c], Vinit Parida [b],[c]

[a] *Department of Operations Management, Copenhagen Business School, Solbjerg Plads 3, 2000, Frederiksberg, Denmark*
[b] *Department of Business Administration, Technology and Social Sciences, Luleå University of Technology, 971 87, Luleå, Sweden*
[c] *Department of Business, History and Social Sciences, USN School of Business, University of South-Eastern Norway, Campus Vestfold, Norway*

ABSTRACT

Industrial manufacturers increasingly develop digital platforms in the business-to-business (B2B) context. This emergent form of digital platforms requires a profound yet little understood holistic perspective that encompasses the co-evolution of platform architecture, platform services, and platform governance. To address this research gap, our study examines multiple platform sponsors from an industrial manufacturing context. The study demarcates three platform archetypes: product platform, supply chain platform, and platform ecosystem. We argue that each platform archetype involves a gradual development of platform architecture, platform services, and platform governance, which mirror each other. We also find that each platform archetype is characterized by a specific innovation mechanism that contributes to the platform service discovery and expands the platform value. Our study extends the co-evolution perspective of platform ecosystem literature and digital servitization literature.

## 1. Introduction

Digital transformation profoundly changed the way firms innovate and secure competitiveness (Hanelt et al., 2020; Nambisan et al., 2019; Yoo et al., 2012). For instance, digital technologies such as the Industrial Internet of Things (IIoT) have allowed firms to make products and services smarter (Kiel et al., 2017; Porter and Heppelmann, 2014; Raff et al., 2020), blur the lines between physical products and digital services (Hanelt et al., 2020; Jocevski, 2020) and unlock vast innovation opportunities (Lanzolla et al., 2020). Yet, to reach the full potential of digital transformation, firms need to adopt digital platforms (Parker et al., 2016; Sandebrg et al., 2020). Digital platform firms use digital technologies to "exploit and control digitized resources that reside beyond the scope of the firm, creating value by facilitating connections across multiple sides, subject to cross-side network effects" (Gawer, 2020).

However, firms approach digital platform development in different ways (Cennamo et al., 2020). While some firms use digital technologies to build multi-sided platforms (e.g., Spotify, Netflix, Uber, Airbnb), others use it to collaboratively expand the platform value with their customers, suppliers, technology providers, and competitors (e.g., Volvo, Komatsu, BMW) (Adner et al., 2019; Cennamo et al., 2020). In particular, industrial firms in the business-to-business (B2B) setting have also started embracing so-called *digital servitization.* Literature defines it as a large-scale transformation in processes, capabilities, and offerings within industrial firms and their associated ecosystems, to progressively create, deliver, and capture increased service value, arising from a broad range of enabling digital technologies (Sjödin et al., 2020b). For example, smart and connected products, combined with artificial intelligence (AI) capabilities, have enabled manufacturers, such as Volvo and Komatsu, to develop an ecosystem that brings together different vehicle markets, connectivity providers, applications, and customers (Raff et al., 2020).

Digital servitization is enabled by connecting installed bases of industrial assets and equipment to an industrial digital platform that provides aggregation of data and analytical capabilities for greater value creation and capture (Kiel et al., 2017; Paiola and Gebauer, 2020). Thus, previous research showed that digital servitization goes hand-in-hand with a *platform approach* (Cenamor et al., 2017; Leminen et al., 2020; Rajala et al., 2019; Wei et al., 2019). For instance, industrial digital platforms allow to connect various IIoT-enabled machines, collect operational and equipment data, and conduct cutting-edge analytics to provide advanced platform services, such as preventive maintenance, fleet management, or even site optimization (Kohtamäki et al., 2020).

Industrial platform sponsors gradually open up their platforms to a specific set of complementors, such as technology providers or traditional intermediaries (Hilbolling et al., 2020), in order to develop more advanced platform services. This creates opportunities for higher value creation and capture through an *ecosystems approach* (Jacobides et al., 2018; Kretschmer et al., 2020). However, such advanced use of industrial digital platforms is mostly in a nascent stage of development and further inquiry is needed.

Based on a review of studies, a *platform ecosystem* can be viewed as an evolving meta-organizational form characterized by enabling *platform architecture*, supported by a set of *platform governance* mechanisms necessary to cooperate, coordinate and integrate a diverse set of organizations, actors, activities, and interfaces, resulting in an increased *platform value* for customers through customized *platform services* (Cennamo, 2019; Constantinides et al., 2018; Hou and Shi, 2020; Jacobides et al., 2018; Kretschmer et al., 2020). The mirroring hypothesis suggests that a technological architecture and associated governance mechanisms "mirror" one another in the sense that the structure of one will correspond to the structure of the other (Colfer and Baldwin, 2016). Despite this, it is only recently that the literature has started exploring the co-evolution of platform architecture, platform services, and platform governance (Hou and Shi, 2020; Rietveld and Schilling, 2020; Saadatmand et al., 2019; Tiwana et al., 2010). Many research gaps remain open for investigation related to digital servitization and industrial digital platforms.

First, a major concern regarding the current platform research is that scholars largely exclude the design of platform architecture (Cennamo, 2019; Thomas et al., 2014; Tiwana, 2014) and associated features of the value proposition (Kiel et al., 2017; Leminen et al., 2020; Schroeder et al., 2020), from the unit of analysis. For instance, to reach the higher-order platform services such as optimization and autonomous services, a platform sponsor needs to add a complementary set of external modules to the platform core (Constantinides et al., 2018). Moreover, a platform sponsor needs to make careful architectural decisions to support the platform growth (Cennamo and Santaló, 2019; Tilson et al., 2010). Thus, the interplay between the platform architecture development and platform services is important to take a more progressive approach towards the full-scale digital servitization transformation.

Second, the current platform literature lacks a process perspective in terms of how the platform architecture development and associated platform governance mechanisms co-evolve. Scholars predominantly focus on the B2C digital marketplaces at advanced stages with an established pool of complementors and customers (Cennamo, 2019). In the B2B context, a platform sponsor usually develops a proprietary platform with an exclusive set of complementors and customers (Eisenmann, 2008; Rietveld et al., 2019) and gradually opens up to other complementors (Broekhuizen et al., 2021; Cenamor and Frishammar, 2021; Wei et al., 2019). Thus, the emergence of a platform ecosystem in the B2B context gradually unfolds through close collaboration between a platform sponsor, complementors, and customers (Enkel et al., 2020; Granstrand and Holgersson, 2020) as well as competing platforms (Adner et al., 2019). Moreover, functional contributions of actor-specific data are vital in the emergence of platform governance as they actively shape the platform growth (Alaimo et al., 2020). Therefore, there is a need to further understand the co-evolution of platform architecture and platform governance in the B2B context.

Third, scholars argue that digital generativity challenged the traditional assumptions of value creation and capture and raised the need for new theory development (Nambisan et al., 2019; Yoo et al., 2010). Advances in artificial intelligence (AI) and the growing availability of data that cut across diverse platform ecosystem members (Alaimo et al., 2020; Haefner et al., 2021), allow rapid scaling of knowledge search and more effective knowledge recombination (Lanzolla et al., 2020; Lenka et al., 2017; Savino et al., 2017). This results in new platform services that increase the platform value for customers (Cennamo, 2019; Gregory et al., 2020). However, it is only recently that studies started exploring the underpinnings of the platform service discovery (Dattée et al., 2018; Hou and Shi, 2020). Consequently, empirical insights from digital servitization can be helpful in further understanding the innovation potential arising from a platform ecosystem.

Considering the aforementioned gaps in the literature, the evolution of industrial digital platforms requires more attention (Hou and Shi, 2020; Nambisan et al., 2019). More specifically, the literature lacks a process perspective on how the co-evolution of platform architecture, platform services, and platform governance expands the platform value. In this paper, we develop new theoretical insights into the evolution of industrial digital platforms as a vehicle for the digital transformation of manufacturing firms. More specifically, the purpose of this paper is *to investigate how industrial manufacturers can expand the platform value through the evolution of industrial digital platforms.*

Our study draws on rich data from four world-leading construction equipment manufacturers. These platform sponsors have gradually invested in the platform architecture that supports highly advanced platform services. Our findings delineate three phases in the evolution of platform architecture: (1) product data collection, (2) analytics utilization, and (3) artificial intelligence enablement. We show that each platform architecture phase links with the specific platform governance – namely, (1) value chain expansion, (2) value system expansion, and (3) ecosystem expansion. Hence, both platform architecture and platform governance need to be considered simultaneously, in each phase of the industrial digital platform evolution, to expand the platform value through (1) monitoring service development, (2) optimization service development, and (3) autonomous service development. We also find that each phase is characterized by a specific innovation mechanism: (1) search depth, (2) search breadth, and (3) recombination. Consequently, we demarcate three platform archetypes: (1) product platform, (2) supply chain platform, and (3) platform ecosystem. Our study extends the co-evolution perspective of platform ecosystem literature and digital servitization literature.

## 2. Theoretical background

### 2.1. Digital servitization and industrial digital platforms

Digitalization is increasingly regarded as one of the most prominent drivers of innovation (Nambisan et al., 2019). Manufacturing firms have been at the forefront of the adoption of advanced wireless sensors that provide information about the environment, context, and location of their industrial assets (Leminen et al., 2020; Ng and Wakenshaw, 2017; Zheng et al., 2020). These technologies allowed digital components and electronics to be embedded in various "things" and enabled them to be smarter and progressively more interlinked – often called the Industrial Internet of things (IIoT) (Porter and Heppelmann, 2014; Sestino et al., 2020; Suppatvech et al., 2019). Today, manufacturers link physical and digital worlds and collect data from diverse objects, devices, and machines (Bilgeri et al., 2019; Leminen et al., 2018; Schroeder et al., 2020). As a result, they are able to make new digital products that perform better in terms of productivity and profitability (McAfee and Brynjolfsson, 2012).

Recent studies view digitalization of manufacturing as closely related to servitization transformation (Kohtamäki et al., 2020). In particular, the emerging literature on digital servitization shows that digital technologies allow manufacturers to provide new value creation and value capture opportunities through monitoring, control, optimization, and autonomous functions (Gebauer et al., 2020; Kohtamäki et al., 2019; Paschou et al., 2020). In other words, digital technologies not only facilitate servitized business models (Arnold et al., 2016; Suppatvech et al., 2019; Vendrell-Herrero et al., 2017), such as outcome business models (Sjödin et al., 2020a; Visnjic et al., 2017) but also contribute to the development of emerging industrial digital platforms with opportunities for value creation and capture that go far beyond traditional products and services (Cenamor et al., 2017; Rajala et al., 2019; Wei et al., 2019). However, the industrial digital platform development has implications for industrial manufacturers. First, manufacturers are required to mature in terms of platform architecture and shift away from firm-based product platforms to platform ecosystems (Sandebrg et al., 2020). Second, they are required to manage a wide range of tensions that relate to platform governance (Rietveld and Schilling, 2020).

*2.2. The evolution of platform architecture*

A platform ecosystem is an evolving meta-organizational form where the platform architecture is a shared technological core that supports the ecosystem's members to create and capture value (Hou and Shi, 2020; Kretschmer et al., 2020). A platform ecosystem is usually organized around a hub firm that owns or sponsors the platform (Rietveld and Schilling, 2020). A platform sponsor designs the platform architecture that describes how a relatively stable platform core, with specific design rules and a diverse set of complementary modules, allows stakeholders to orchestrate data collection, data storage, data flow, data aggregation, and data commercialization (Alaimo et al., 2020; Constantinides et al., 2018; Tiwana et al., 2010). However, a key concern for manufacturers that operate in the B2B context is how the platform architecture can evolve and extend its functional scope to effectively serve emerging opportunities for future services (Agarwal and Tiwana, 2015; Koutsikouri et al., 2018; Tilson et al., 2010)?

In the initial phase of industrial digital platform development, manufacturers often develop product platforms around key installed bases (Gawer, 2014; Sandebrg et al., 2020; Svahn et al., 2017). Product platforms incorporate digital modules that are programmable, addressable, sensible, communicable, memorable, traceable, and associable (Warner and Wäger, 2019; Yoo et al., 2010) and allow separation of form from the function (Autio et al., 2018). Therefore, the integration of digital modules allows manufacturers to collect valuable data on the installed bases (Björkdahl, 2020) and render new functionalities such as monitoring or visualization services that generate supplementary revenue streams (Zhu and Furr, 2016).

Next, to reach the higher-order platform services such as optimization and autonomous services, a platform sponsor needs to add a complementary set of external modules to the platform core (Constantinides et al., 2018). External modules may come in form of advanced sensors, data analytics, applications, or cloud-based data storage (Iansiti and Lakhani, 2020). Consequently, the effectiveness of such platform architecture depends on the management of different modules that are introduced over many years, for different purposes, and by diverse actors (Henfridsson and Bygstad, 2013). More specifically, for manufacturers, a large part of the technological setup and digital knowledge may lie outside their organizational boundaries, which contributes to greater dependency on the surrounding digital ecosystem (Björkdahl, 2020; Hanelt et al., 2020). For instance, external modules may come from traditional suppliers and distributers to start-ups and tech-giants (Hanelt et al., 2020; Visnjic et al., 2018). As a result, it is difficult for manufacturers to fully control the platform architecture (Eisenmann, 2008). It also requires from manufacturers to reconsider platform boundary choices (Gawer, 2020; Huikkola et al., 2020) and increase platform openness (Cenamor and Frishammar, 2021).

The platform literature shows the utility of boundary resources in the platform growth (e.g., API) (Ghazawneh and Henfridsson, 2013). For instance, boundary resources are 'resourceful' as their design enables external contributions from heterogeneous actors that increase the platform value (Cennamo, 2019; Yoo et al., 2010). Boundary resources can also be 'securing' as they prevent the development of applications that risk to damage the overall platform value (Ghazawneh and Henfridsson, 2013). Thus, architectural control decisions on boundary resources allow to balance between promoting and constraining the platform growth (Cennamo and Santaló, 2019; Tilson et al., 2010).

*2.3. The evolution of platform governance*

Platform governance requires addressing tensions related to platform openness and control but also managing simultaneous collaboration and competition with complementors (Rietveld and Schilling, 2020). Industrial digital platforms focus on the B2B context that is largely understudied in the platform ecosystem literature and presents a relevant empirical context for the investigation.

First, in the B2B context, actors actively shape the platform ecosystem through bilateral governance mechanisms between the platform sponsor and prospective members. In other words, the platform sponsor does not draw the governance structure for other actors to fill in their roles and positions (cf. Adner, 2017). As a result, the platform sponsor has less control over the evolution of the platform ecosystem (Gawer and Henderson, 2007). Therefore, the ecosystem governance is emerging, which presents a more nuanced setup to study strategic interactions between the platform sponsor, complementors, and customers (Panico and Cennamo, 2020).

Second, the majority of industrial digital platforms emerge as proprietary platforms (Eisenmann, 2008), where the platform sponsor usually initiate the platform development with an exclusive set of complementors (Hilbolling et al., 2020), traditional intermediaries (Randhawa et al., 2018), and customers (Sjödin et al., 2020b), what resembles a supply chain platform logic (Gawer, 2009). In contrast to the B2C digital marketplaces, complementors cannot join based on self-selection (cf. Gulati et al., 2012). Subsequently, the platforms sponsor gradually opens the industrial digital platform through the selective promotion of complementors (Rietveld et al., 2019). However, the platform sponsor needs to make careful strategic decisions about how many and what type of complementors it wants to induce to join the platform (Rietveld and Schilling, 2020). Current literature lacks clarity about the antecedents of complementor selection in the evolution of platform ecosystems (Hou and Shi, 2020; McIntyre and Srinivasan, 2017). Moreover, the industrial B2B context may shed light on governance mechanisms in the early phase of platform ecosystems (Hannah and Eisenhardt, 2018).

Finally, due to the complexity around the onboarding prospective ecosystem members, platform size and platform scope are rather narrow in the early phase of the industrial digital platform development (Cennamo, 2019; Gawer, 2020). The platform sponsor usually focuses on a specific profit foci around its industrial assets and equipment, which creates opportunities for complementors and competing platforms with different profit foci to cooperate and capture more value when they interact through the specific platform (Adner et al., 2019). Yet, the current literature is underexplored in terms of platform strategic positioning (Pellizzoni et al., 2019) and platform coopetition in the industrial B2B context (cf. Basaure et al., 2020). Finally, such context may provide a useful ground to study platform competition that goes beyond the winner-takes-all approach (Mcintyre, 2019).

*2.4. The platform service discovery and the platform value*

Platform services bring together the platform architecture and platform ecosystem's members through joint exploitation activities (Jacobides et al., 2018; Tilson et al., 2010). However, the current ecosystem literature has diverging views about how the focal value proposition, that is, platform service, came about (Hou and Shi, 2020). While some authors reason that the value proposition defines the configuration of ecosystem activities (Adner, 2017), others see value proposition as a derivate of modular structures (Jacobides et al., 2018). Recently, Hou and Shi (2020) point in the direction that the value proposition is a result of the co-evolution of ecosystem members that are embedded in micro and macro contexts. Still, the literature presents ambiguity around the antecedents of the value proposition discovery (Dattée et al., 2018).

Platform services have been frequently associated with greater possibilities for innovation (Nambisan et al., 2017). Moreover, platform services leverage digital technologies that allow rapid scaling of knowledge search as well as more effective knowledge recombination (Lanzolla et al., 2020; Lenka et al., 2017; Savino et al., 2017). As a result, new platform services can often be developed at a marginal cost (Rifkin, 2015). Therefore, new functionalities expand the overall platform value as they increase benefits that customers can derive from using the platform (Cennamo, 2019).

Empirical insights from industrial digital platforms can be helpful in further understanding the antecedents of the platform service discovery. First, digital platforms support firms to deepen the knowledge search within current knowledge structures (Lanzolla et al., 2020). For instance, sophisticated algorithms can find patterns within existing data sets that were previously not possible to identify, such as in the case of error detection (Kieu et al., 2018). Second, digital platforms can also broaden the knowledge search with novel inputs from various objects, devices, and machines. For example, the broad inclusion of various inputs contributed to the development of advanced feel management systems (Crainic et al., 2009; Haefner et al., 2021). Finally, the inherent modular structure of digital platforms allows for recombination (Autio et al., 2018; Yoo et al., 2010). Weitzman (1998) argues that the greater the number of recombinable modules, the more opportunities for creation of novel solutions. For instance, agile development allows to recombine a multitude of micro-services into complex solutions (Ghezzi and Cavallo, 2020; Sjödin et al., 2020b). Thus, the platform value expands as the synergistic interaction of micro-services increase platform functionalities (Bozan et al., 2021; Tiwana et al., 2010).

## 3. Methodology

We conducted the study using four in-depth cases of manufacturers that gradually developed industrial digital platforms around their industrial assets and equipment. We opted for a qualitative study with a grounded theory building approach (Glaser and Strauss, 1967) as it responds better to "how" questions than quantitative research's input–output models (Yin, 2017). Moreover, our multiple case studies generated rich, field-based insights into the evolution of industrial digital platforms.

### 3.1. Research setting and sample

In order to investigate how platform sponsors approach the development of industrial digital platforms, we adopted an inductive case study design. The present study is based on four global equipment manufacturers (hereafter Alpha, Beta, Gamma, and Delta). Multiple case studies allow to generate multiple observations on complex and simultaneous processes (Eisenhardt and Graebner, 2007; Gioia et al., 2013) and to develop detailed insights of the theoretically novel phenomenon (Edmondson and Mcmanus, 2007). This research design was particularly useful since there is limited knowledge about the evolution of industrial digital platforms. Information from real-life cases can help identify new aspects and factors derived from reality (Yin, 2017).

The selected manufacturers are world-leading providers of equipment for construction and related industries. They have taken significant steps to restructure their organizations and processes to ensure successful digital platform development. Moreover, they also engaged in collaboration with various external partners to develop and deliver advanced platform services. Currently, these digital platforms already offer various platform services, including geo-location monitoring services, proactive failure detection, and fuel consumption, but also advanced platform services such as fleet management services and agreed-upon availability services with risk-reward sharing. Moreover, all digital platforms reached a level of technological maturity where platform sponsors started developing AI-driven autonomous solutions (e.g., co-pilot services) together with complementors and industry-level competitors. Finally, we had established good contacts with each manufacturer, which led to the collection of detailed descriptions of the digital platform's development trajectory and in-depth information about the process and key milestones.

### 3.2. Data collection

Between May 2018 and January 2020, the authors gathered data primarily through two sources: cross-functional interviews and archival data (Yin, 2017). A semi-structured interview guide was developed that aimed to unfold the process of digital platform development (Fontana and Frey, 1998). In total, the authors conducted 48 interviews with key informants. The informants were identified by snowball sampling, where key informants such as vice presidents were asked to recommend people who had an active role in different phases of digital platform development (Kvale, 1996). Informants who had both functional and senior roles were interviewed in order to capture a multifaceted view of the process. This approach was deemed necessary because the evolution of digital platforms typically requires complex interactions between multiple organizational functions. The informants included senior vice presidents, chief technology officers, business development managers, R&D managers, project managers, key account managers, product managers, portfolio managers, and other managers. These informants gave us a wider understanding of the case studies. We concluded the data collection when theoretical saturation was reached (Glaser and Strauss, 1967). Table 1 summarizes the case studies and the data collection efforts.

With the support of an interview guide, the informants were asked open-ended questions. The interview guide targeted themes about digital servitization, technological architecture development, partner engagement, and how platform sponsors developed a portfolio of platform services. For example, informants were asked questions such as: How did you initiate the platform development process from the technology perspective? What were the initial functionalities of the platform? What platform services you initially offered to customers? How did platform architecture relate to platform service development? How and when did you involve partners in the platform development process? In seeking answers to these overarching questions, we encouraged informants to base their answers not only on the current status of their digital platforms but also on the initial steps taken that preceded the emergence of the platform ecosystems, so that the process could be captured. Follow-up questions were used to clarify points and obtain further details, which enabled further exploration of relevant steps in the development process. The interviews took approximately 60–120 min each and were held face-to-face or via online conference calls.

We combined retrospective data and real-time data (Miller et al., 1997). In order to mitigate retrospective bias we focused on concrete events in the platform development process (Miller and Salkind, 2002). Moreover, we used archival data to track changes in the platform development process. We performed document studies, reviewing company reports, presentations, and newspapers to validate and provide context to our informants' views. This additional effort helped us triangulate data collected from different sources and increase data reliability (Rowley, 2002). Finally, we recorded and transcribed the interviews, or, in the cases where informants did not wish to have their interviews recorded, took extensive notes during and after the interview (Yin, 2017).

### 3.3. Data analysis

The data analysis was based on a thematic analysis approach, which provides ways to identify patterns in large, complex data sets (Braun and Clarke, 2006). Moreover, the thematic analysis offers means to effectively and accurately identify empirical themes that are grounded in the case study context. Through a series of iterations and comparisons (Alvesson, 2011), the authors grouped empirical themes into the conceptual categories. Moreover, the conceptual categories reflected the theoretical constructs from the literature (Strauss and Corbin, 2015). Next, the authors performed a cross-case analysis (Eisenhardt, 1989) to identify similar empirical themes and categories across the cases (Miles and Huberman, 1994). Finally, the conceptual categories were clustered into the aggregate dimensions (Gioia et al., 2013). Therefore, we followed a three-step process similar to that described in the literature (Braun and Clarke, 2006; Gioia et al., 2013).

**Table 1**
Descriptions of case studies and data collection efforts.

| Firm | Alpha | Beta | Gamma | Delta |
| --- | --- | --- | --- | --- |
| *Revenue* | $ 4120 M | $ 23.1 B | $ 22.5 B | $ 10.389 B |
| *Employees* | 38,000 | 101.500 | 14,000 | 41,670 |
| *Main products and services* | The company is the world's leading manufacturer of equipment for construction, mining, and agriculture. | The company is one of the world's largest construction equipment manufacturers. | The company is the world's leading manufacturer of specialized equipment for drilling and rock excavation, and a complete range of related consumables and services. | The company is the world's leading supplier of equipment, tools, services and technical solutions for the mining and construction industries. |
| *Data collection* | Semi-structured interviews | Semi-structured interviews | Semi-structured interviews | Semi-structured interviews |
|  | Internal and external documentation | Internal and external documentation | Internal and external documentation | Internal and external documentation |
|  | Site visits | | Workshops | Workshops |
|  | Workshops | | Industry presentations | Industry presentations |
| *Interviewee role(s)* | Executive (1), Product portfolio manager (2), Project manager (3), R&D manager (2), Senior project manager (1), Procurement manager (1), service delivery manager (3), regional manager (2) | Vice president (1), Director (3), Service manager (3), Division manager (1), Strategy manager (1), Digital and technology director (1) | Senior manager (3), Procurement manager (1), IT manager (2), Head of automation (1), key account manager (3), business development manager (2), project manager (1) | Executive (2), Product manager (2), Line manager (1), IT manager (2), Project manager (1), automation lead (1), automation and digital solution manager (1) |
| *Total interviews* | 15 | 10 | 13 | 10 |
| *Industrial digital platform* | *The industrial digital platform features fleet management, performance management, and jobsite solutions that increase safety, productivity, and uptime.* | *The industrial digital platform features solutions for individual machines, whole fleets or entire jobsite, including autonomous and semi-autonomous solutions.* | *The industrial digital platform features solutions for increasing fleet efficiency, safety, and optimizing fleet performance in real-time.* | *The industrial digital platform features solutions for increasing security, reliability, and performance of equipment through a seamless integration of various digital solutions.* |

The first step in our data analysis was an in-depth analysis of the raw data (i.e., the interview transcripts). This analysis consisted of reading interview transcripts, highlighting phrases and passages. By coding the common words, phrases, and terms mentioned by informants, we identified empirical themes that reflect the views of the informants in their own words. The second step of the analysis was to further examine the empirical themes to detect links and patterns among them. This iterative process yielded conceptual categories that represent theoretically distinct concepts created by combining empirical themes. The third step involved the aggregation of conceptual categories. Here, we used insights from the literature to form theoretically-rooted dimensions. More specifically, the authors grouped the conceptual categories into three aggregate dimensions: The first dimension *platform architecture* unites the categories "product data collection," "analytics utilization," and "artificial intelligent utilization". The second dimension *platform services* contain the categories "monitoring service development," "optimization service development," and "autonomous service development". Finally, the dimension *platform governance* consists of the categories "value chain expansion," "value system expansion," and "ecosystem expansion". This step of the data analysis involved thorough discussions about the data structure. Internal validity tests were conducted to ensure greater accuracy of the data structure through email correspondence and follow-up discussions with selected informants.

As the last step, we analyzed the logic, linkages, and mirroring across aggregate dimensions, conceptual categories, and empirical themes. We were able to demarcate three phases in the industrial digital platform evolution that we label platform archetypes: "*product platform,*" "*supply chain platform,*" and "*platform ecosystem*". Moreover, we wanted to explicate the antecedence of the platform service discovery for each archetype. We used insights from the literature to identify conceptual categories related to innovation mechanisms "search breadth," "search depth," and "recombination". This practice allowed us to generate an overall model (see Fig. 2) while Fig. 1 shows the entire data structure that resulted from the data analysis. Table 2 provides examples of illustrative quotations for the conceptual categories. The initial results of the study were presented to 10 key informants from case studies to validate the results through discussion. Further adaptations were made where relevant.

## 4. The evolution of industrial digital platforms

In this section, we present a holistic model of the industrial digital platform evolution that emerged based on the analysis of the four cases. We present our findings in three parts, each corresponding to one of the key dimensions in the industrial digital platform evolution: platform architecture, platform services, and platform governance. Following the presentation of the findings, we offer the resulting framework where we further elaborate on three platform archetypes and the underlying innovation mechanism for each archetype.

### 4.1. Platform architecture

There is no one-size-fits-all approach to the industrial digital platform development. However, this study shows that a key part of the digital transformation journey was investing in the technology of the platform core. During the initial phase, platform sponsors tended to invest in the platform architecture progressively and increase the capacity for **product data collection**. This included enabling data gathering for major installed bases ex-ante to the possible use cases. Next, platform sponsors focused on **analytics utilization** as advanced sensors provided increased data quality and data variety. It enabled platform sponsors to start aggregating data, correlating different data sets, and finding patterns. Finally, **artificial intelligence enablement** brought the power of AI and platform openness that could leverage external data sources and reveal hidden insights. Overall, a key initial milestone in the platform architecture development was investing in the sensor network that generated data and allowed a higher degree of connectedness within the industrial assets. Henceforth, data aggregation and data analytics unlocked opportunities for higher value creation through collaboration with external partners.

**Product data collection** refers to efforts in gathering and sharing

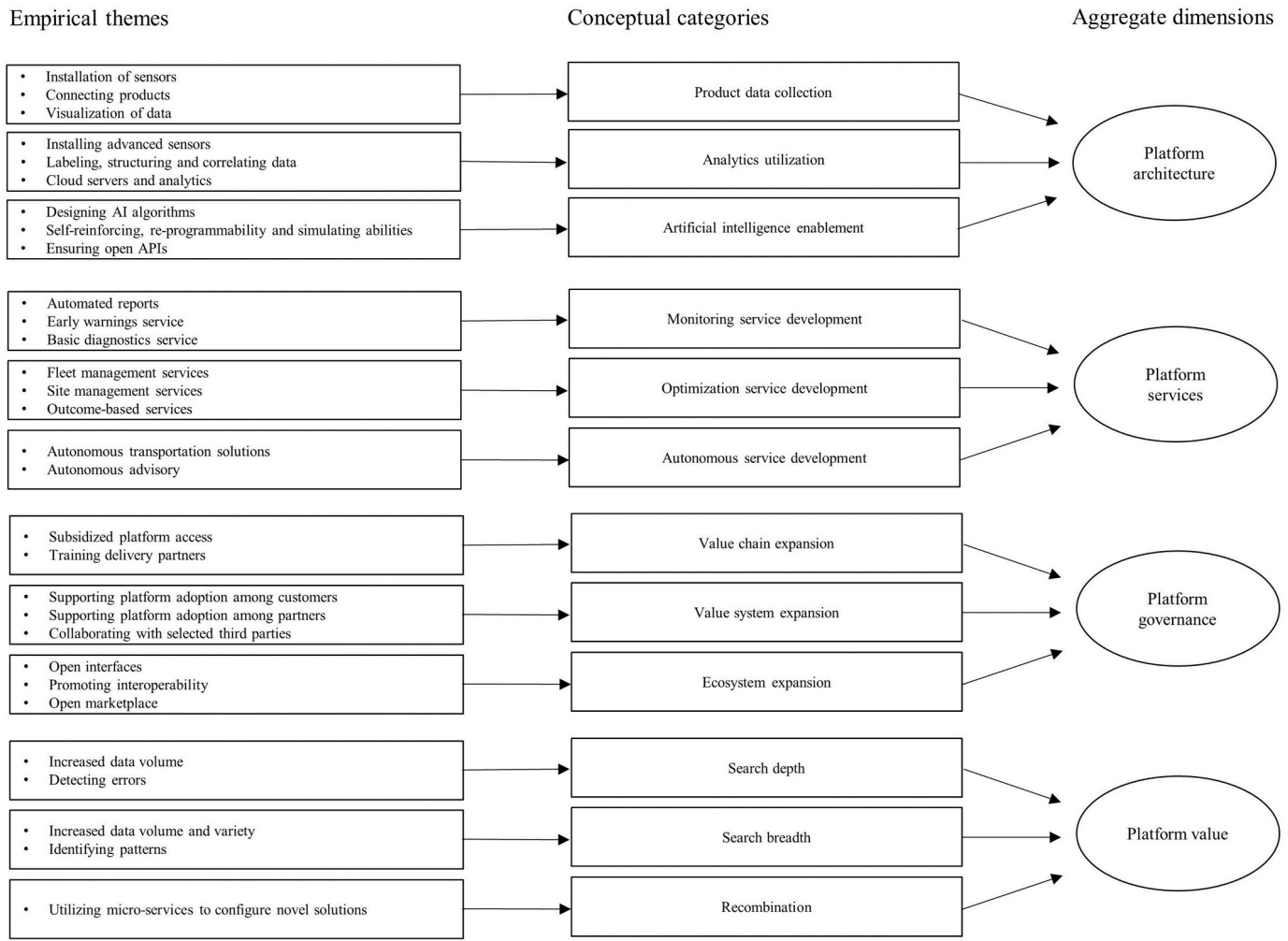

Fig. 1. The data structure.

data about product portfolios, creating a wealth of information about equipment's daily usage, user behavior, and faults. Initially, informants reported that the *installation of sensors* provided a good starting point. Digital sensor data helped manufacturers to have a better understanding of the product in use, gather operational insights, and look for potential inefficiencies. For example, platform sponsors initially installed sensors for measuring oil quality and temperature, which facilitated automatic measurement (i.e., digital). However, installing sensors was not enough; a critical point was ensuring that sensors allowed *connecting products* to the network and remote monitoring of various aspects of the product performance. The installation of sensors was initially a machine-centric exercise that offered limited opportunities for creating new services. While the sensor data unlocked advanced remote monitoring services, *visualization of data* was found to be extremely important for customers. As a result, platform sponsors created digital portals – an embryonic version of the digital platform. For example, one executive explained how they installed a series of sensors focused on connecting as many machines as possible. He elaborated:

*We started 2001 by connecting all our biggest machines, with no idea what we were going to use the data for, to be frank. It was more of a strategic decision, saying that 'we are probably going to use the data in the future to build services'. That's where it all started, then the current CEO took the decision to accept that cost, so today we have thousands of machines connected.*

*Then, we started looking at creating some intelligence in this. We created a portal called the (portal name), which consists of the basic information for the individual machines. So, we could provide a very machine-centric view to* *our customers. We took one machine and we showed the fuel efficiency, the idle time, and basic machine data and visualized this for the customer in the portal.*

The next phase included **analytic utilization** as platform sponsors began *installing advanced sensors* that were able to collect data about geo-location, load measurement, hydraulic assessment, etc. One informant highlighted that there are approximately 400 sensors per machine that funnel the data to the platform. Such data sets allowed to proactively discover anomalies so that appropriate actions can be taken in advance, if necessary, in terms of both safety and efficiency. As data volume and data variety rapidly increased, a critical investment was needed to ensure that all collected data is utilized. This required *labeling, structuring, and correlating* data that previously resided in silos. Having structured databases made advanced analytics more feasible and allowed a larger number of users (i.e., data scientists and engineers) to understand and analyze the data. Moreover, these advanced sensors allowed machines to communicate with each other, with the infrastructure, and created significant machine-to-server activity. Thus, the real-time visibility of data was required. The *cloud servers and analytics* enabled the platform sponsors to access the data from any location and, more importantly, transformed the way platform sponsors use the data and develop services, which shifted from operational to strategic level (e.g., headquarters). Accordingly, cloud servers made possible to display live analytics dashboards to customer's executives. A service manager elaborated:

*Increasingly, all the important data are being stored and analyzed in our proprietary cloud platform. This shift has been revolutionary for our*

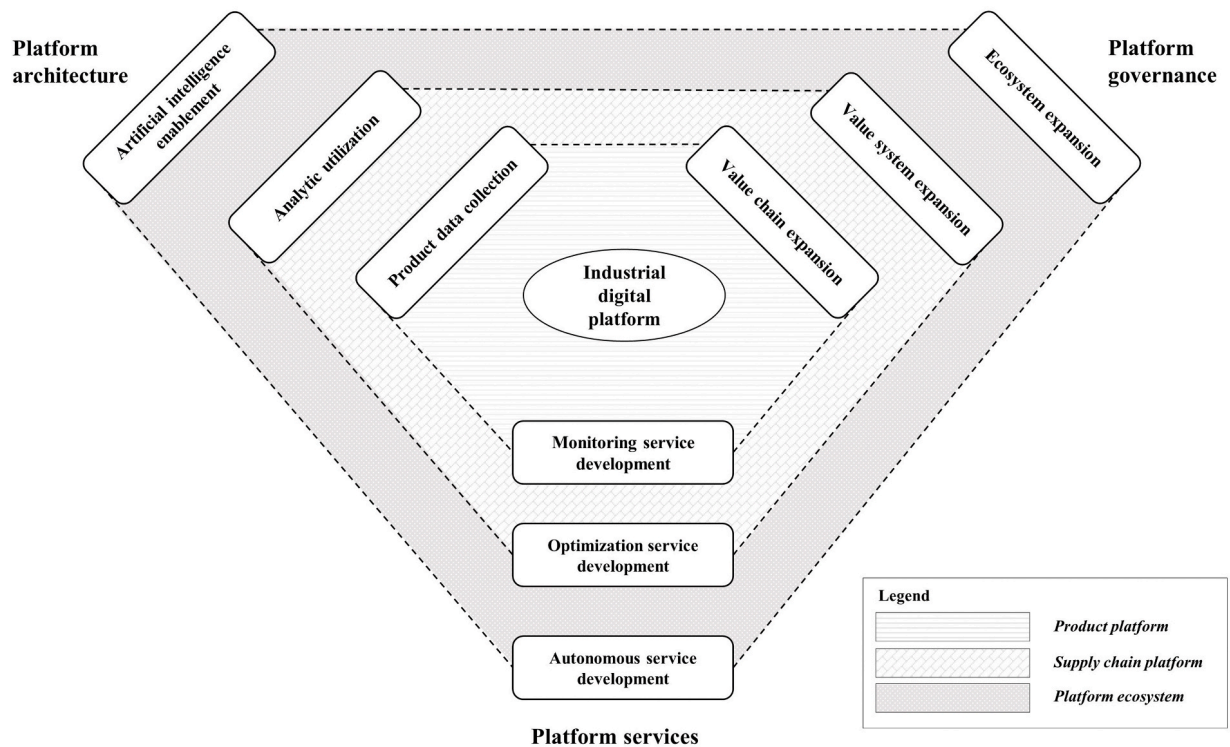

Fig. 2. The evolution of industrial digital platforms.

*distributed experts as they can provide closer to real-time data insights to the service engineers and customers.*

Finally, the most advanced platform architecture included **artificial intelligence enablement**. Platforms sponsors started including the lidar (i.e., optical radar) data that created a detailed picture of a machine's surroundings. In such scenarios, machines can understand the structure of their workspace and the objects that surround them. Artificial intelligence algorithms analyzed a combination of various sensor streams to enable machines to act autonomously. First, *designing AI algorithms* was the key activity on the way to full autonomy. This initial design of AI algorithms aimed at finding the root-cause problem across various functions. For instance, potholes can contribute to tire aging and deterioration. However, if the algorithm navigates the machine, it will bypass potholes and the tire's lifespan will increase. Similarly, controlled acceleration and deceleration of vehicles can enormously reduce fuel costs. Moreover, autonomous machines can operate in complete darkness, which lowers operating costs and eliminates on-site deaths caused by object strikes. Second, the real power of algorithms is their ability to learn. More specifically, they provide *self-reinforcing, re-programming, and simulating* abilities. For instance, the algorithmic ability to analyze data at high speed and accumulate knowledge allowed to pre-program new machines with the intelligence of all other machines that preceded them. One executive summed it up:

*We started looking more into machine learning and analytics to add to proactive monitoring, but also other areas such as spare part logistics, simulations, and operator assistance functionalities. It's low scale but we look into different opportunities.*

*The value of building on machine learning is that we can build on data insights from all our active and connected machines, not only the ones that customers own. For example, the bucket filling function in the mining environment. This, combined with learning based on a unique customer operational environment, allows for a significantly higher value.*

Finally, the ability to improve algorithms depends on the inflow of high-quality data, which requires more interconnected and open systems. Platform sponsors started to collaboratively develop advanced platform services through *open APIs*. Platform sponsors started opening up their interfaces in an authorized way to extend the platform value and meet customers' requests. Platform openness offered vast opportunities for creating value that goes beyond manufacturers' core competencies. For instance, a collaboration with specialized external partners such as drone surveying and 3D topology technology partners unlocked the benefits for their mutual customers. A product portfolio manager highlighted the importance of opening up platform architecture to complementary partners:

*We already have API going out. We also have third-party applications that we work with, but in limited numbers. We do strategic partnerships.*

*Initially, it has been a challenge to have a totally open platform, as we need to ensure adherence to safety regulations and legal requirements. But it has always been our ambition to have an open platform where other suppliers can develop applications.*

*It's not possible to have the leading digital platform in the industry using only in-house development of applications. We have to be progressive and allow other technology companies to gain from our data and deliver high value to our customers. This is a win-win situation for us, our customers, and other actors.*

### 4.2. Platform services

Platform sponsors progressively developed more advanced platform services building on the platform architecture functionalities. Our analysis uncovered that platform service development closely mirrored three phases of the platform architecture development. Therefore, we delineated three levels of platform services: **monitoring service development**, **optimization service development**, and **autonomous service development**.

Platform sponsors began with **monitoring service development**, which initially took a machine-centric view and focused on creating *automated reports* (e.g., fuel analysis). Connected machines unlocked the possibility of providing active remote monitoring and generating automated reports. Automated reports served as a base of all other more advanced platform services. Next, *early warning services* presented a natural extension to automated reports and an opportunity for

**Table 2**
Representative quotes.

| Dimensions and categories | Representative quotations |
| --- | --- |
| **Platform architecture** | |
| Product data collection | *We started 2001 by connecting all our biggest machines, with no idea what we were going to use the data for, to be frank. It was more of a strategic decision, saying that 'we are probably going to use the data in the future to build services'. That's where it all started, then the current CEO took the decision to accept that cost, so today we have thousands of machines connected …* |
| Analytics utilization | *Increasingly, all the important data are being stored and analyzed in our proprietary cloud platform. This shift has been revolutionary for our distributed experts as they can provide closer to real-time data insights to the service engineers and customers.* |
| Artificial intelligence enablement | *We already have API going out. We also have third-party applications that we work with, but in limited numbers. We do strategic partnerships.*<br>*Initially, it has been a challenge to have a totally open platform, as we need to ensure adherence to safety regulations and legal requirements. But it has always been our ambition to have an open platform where other suppliers can develop applications.*<br>*It's not possible to have the leading digital platform in the industry using only in-house development of applications. We have to be progressive and allow other technology companies to gain from our data and deliver high value to our customers. This is a win-win situation for us, our customers, and other actors.* |
| **Platform services** | |
| Monitoring service development | *We have machine services; we have data coming from the individual machines where we can look at various efficiency services, such as the fuel consumption, load assist and those things. On the machine level, we have uptime and here we have different things, for example, proactive monitoring. It's not for the whole fleet, it's machine by machine.* |
| Optimization service development | *Then you have fleet services and the customers' needs are very different from those at the machine level. Efficiency services are about fuel efficiency in your fleet. Here, you'd like to be able to analyze maybe 200 machines at the same time and single out the ones that are above the average, for example, when it comes to fuel consumption and understand what application they are so that you could categorize them over time.* |
| Autonomous service development | *If you look at what we did before, the customer paid for the machines and they paid when the machines were broken. So, we have the machine portfolio, and we also have these individual services, proactive monitoring, for example, and we add services to that. However, if you look at what we are doing right now, and what we are trying to expand, it is the use-based models and the outcome-based models. In this we have three different aspects. We have the automation solutions, but it is impossible to sell automation solutions without having performance solutions where we help the customer set this up so that they become more efficient. And then we have these fleet capacity solutions. All three are needed to offer full automation solutions.* |
| **Platform governance** | |
| Value chain expansion | *We have been working for a long time on educating dealers about the potential of our (platform name) system and how to use it in their service processes. Some dealers are more progressive in picking this up and others need more support from us. It is still a conservative industry and there can be resistance. But, we realize that we need our dealers on board to deliver increased customer value from our digital investments.* |
| Value system expansion | *During the early years, we interacted with all customers, but we soon realized that we needed to focus on certain customer segments. For example, we had customers that would buy new machines with our digital platform, but turn off most of its features due to misinformation and lower readiness. So, we started focusing on a few progressive customers and we were able to increase their operational productivity. Such success stories helped us to fine-tune our digital offerings and also encouraged other customers to use the platform.* |
| Ecosystem expansion | *We are increasingly thinking like a technological company, such as Apple. They provide the hardware phone but also generate revenue and create customer value through their App Store. We want to create a construction equipment app store for our customers. Such an open marketplace will have our internally developed recommended apps, but also third-party apps. We are committed to this vision of developing an open marketplace, but this is a challenging undertaking.* |
| **Platform value** | |
| Search depth | *With intelligent equipment, we are able to capture "low hanging fruits".* |
| Search breadth | *We started to develop the [service name] with the main feature of generating load weighing data, among others. Then we started to realize that we can actually start looking at productivity data. And, we realized that we have the opportunity to correlate different things in a more valuable way than we did before. So, we started looking at uptime, what we could do there. We realized that we have other systems data that we could correlate. We looked at different types of data: we looked at statistical data based on quality claims and warranty data. We created a set of data that could be used. When the data were correlated with these patterns of claims for specific systems and machines, we could generate proactive notifications. It was not that long ago that we launched it, two years ago, and we are now also looking at improving our platform in different ways, to become more advanced. Also, we are looking at the productivity and load weighing data, how we can use it to correlate that with other data.* |
| Recombination | *It comes down to what we have to do in the future; if we use micro-services to configure the solutions, we can see this as kind of components of solutions. Before, we used to have our machines, our individual services, and if we combine these, we get solutions … but we are moving away from that … we say we have our portfolio of machines, we have our portfolio of individual services (which are mostly related to uptime), but then you have also digital solutions that leverage both of these other portfolios … we sell these digital solutions as an individual entity. It's a modular approach. Our solution portfolio for machines are very diverse and developed. This allows us to bundle many smaller services into a customized solution.* |

additional value creation and capture. It encompassed the key machine performance data including location, machine hours, availability, fuel consumption, and history of repairs. More importantly, these services warned customers about potential upcoming breakdowns or operator misuse based on deviations in the data. One executive reported that this move was the cornerstone of the shift from reactive platform services to proactive platform services. Proactive platform services unlocked higher value for customers through more sophisticated data analysis. Gradually, platform sponsors started extending the proactive approach, which resulted in *basic diagnostic services*. Basic diagnostic services included assisting services and basic proactive monitoring services. However, they focused on individual machine statistics on utilization, such as fuel consumption and uptime. One executive elaborated:

*We have machine services; we have data coming from the individual machines where we can look at various efficiency services, such as the fuel consumption, load assist and those things. On the machine level, we have uptime and here we have different things, for example, proactive monitoring. It's not for the whole fleet, it's machine by machine.*

The second group of platform services refers to **optimization service development** where the scope was extended from an individual machine to an entire fleet. For instance, *fleet management services* provided customers with the analytical support to eliminate or minimize risks associated with vehicle investment, to improve efficiency and productivity of the fleet, and to reduce the overall fleet transportation and staff-related cost. Vehicle tracking technology (e.g., GPS), driver behavior monitoring, and mechanical diagnostics played a key role in the development of fleet management services. For example, Alpha executives described how fuel consumption and productivity data of the entire fleet helped to single out the outliers. Moreover, fleet management services enabled a high degree of fleet utilization and provided customers with effective tools to analyze and compare detailed characteristics of different machines or models. With the help of these services, customers were able to create the optimal 'assemblage' of machines for every assignment. Consequently, fleet management services gave a comprehensive systemic advantage to the customer, from the operator-level to the customer's headquarters level. A service portfolio manager elaborated on the logic behind fleet management services:

*Then you have fleet services and the customers' needs are very different from those at the machine level. Efficiency services are about fuel efficiency in your fleet. Here, you'd like to be able to analyze maybe 200 machines at the*

same time and single out the ones that are above the average, for example, when it comes to fuel consumption and understand what application they are so that you could categorize them over time.

Also, advanced platform architecture enabled *site management services,* which served to optimize the entire customer's site (e.g., a mining jobsite). This required mapping out the traffic flow and aggregating data across all equipment on the site to identify inefficiencies. Naturally, this required connecting equipment owned by partners or competitors to the platform and integrating them in the analysis. Site management services encompassed site simulations and site path planning that navigates the equipment along the shortest path, site safety services, and productivity improvement services. For example, executives described how analyzing the traffic flow of the entire customer's site allowed them to identify that a particular crossing created a bottleneck as fully loaded machines driving on an incline were forced to stop for other traffic. The site management service recommended giving these machines priority what subsequently led to reducing wear on critical components (e.g., gearbox) and lower fuel costs by up to 5%. Typically, a dedicated site management service team was formed to identify similar issues and increase customer productivity. A manager of product planning described how site management services could improve customer performance:

*So, site services are different from the individual machine or the fleet. Here, it's more about creating the routes and making sure you don't have accidents on site. Then, we also have production monitoring. You need to pinpoint which machines you need to pull data from in order to track production. For example, we look at how much they put in the crusher and relate that to fuel efficiency. Finally, we do site simulations and look at what could drive their machines in shorter cycles.*

With all platform functionalities in place, some customers opted for *outcome-based services,* where the platform sponsor guarantees predefined performance and ensures that all downtime is planned. Outcome-based services entail long-term commitments that encouraged platform sponsors to innovate and continuously improve services. On the other hand, customers trusted that they would always get the best advice because the service performance would directly affect the platform sponsor's bottom line. With outcome-based services, platform sponsors entirely took responsibility for the customers' processes. A manager elaborated:

*Customers can see consistently strong results from their investments that positively influence business outcomes, and the service team at Alpha is empowered with a sense of ownership and incentivized to think big, innovate, and outperform in their own right. This enabled radically re-thinking how we create value. For example, instead of 5 large wheel loaders, Alpha could put in 10 smaller ones, which would reduce the risk of productivity loss when one machine is undergoing maintenance.*

*Then you have uptime services, and here our customers don't buy machines, they buy uptime instead, and that's for the whole fleet. They don't care what machines we put in to those production cycles as long as they produce whatever they need to produce and we then guarantee an uptime for those machines.*

*You can all but guarantee they'll be committed to continued optimization to ensure that their solution consistently meets – or exceeds – expectations. When Alpha wins, the customer wins and vice versa.*

The third group of platform services refers to **autonomous service development**. In the case of autonomous services, advanced platform architecture allowed to further improve flexibility, precision, and productivity of unmanned equipment. While still at the nascent stage, autonomous service development started within the transportation function. Based on site path planning capabilities and self-driving technologies, *autonomous transportation solutions* enabled to use unmanned equipment to transport large volumes of goods on pre-defined routes. First, these services changed the customer's cost structure as driver costs disappeared. Second, they enabled fleets of unmanned equipment to work around-the-clock. In addition, the combination of autonomy and electrification made operations more sustainable. One informant elaborated:

*We have automation solutions and here you have the automated e-checks machines and machines without anyone onboard. They use advanced radars and those kinds of systems that can prevent accidents.*

Finally, *autonomous advisory* solutions incorporated outcome-based logic in autonomous solutions. Autonomous advisory solutions provided dynamic adjustments to customer-centric processes to maximize strategic objectives. Informants shared with us that providing autonomous advisory solutions would not be possible without active collaboration with external partners and customers. Moreover, it was necessary to thoroughly explore the application context and control for the factors that might affect the performance ahead of implementing autonomous advisory solutions. Finally, autonomous advisory does not necessarily mean unmanned technology; it could be the increased level of autonomy, increased safety, efficiency, or sustainability of the construction equipment industry as a whole. A senior executive illustrated the prerequisites for offering autonomous solutions:

*If you look at what we did before, the customer paid for the machines and they paid when the machines were broken. So, we have the machine portfolio, and we also have these individual services, proactive monitoring, for example, and we add services to that. However, if you look at what we are doing right now, and what we are trying to expand, it is the use-based models and the outcome-based models. In this we have three different aspects. We have the automation solutions, but it is impossible to sell automation solutions without having performance solutions where we help the customer set this up so that they become more efficient. And then we have these fleet capacity solutions. All three are needed to offer full automation solutions.*

*But, if they are not interested in automation solutions, then of course we can go with performance solutions and help them optimize their operations. We will look at data coming from our machines or competitors, whatever, and we have a digital platform where we can pull that data to analyze their key machines and the processes that we want to monitor.*

### 4.3. Platform governance

The industrial business-to-business context characterizes a low appetite for risk and requires a high level of privacy. Therefore, platform sponsors gradually induced partners on the supply-side, followed by platform adoption on the demand-side (e.g., customers). The first phase included **value chain expansion**, which implied training, testing, and promoting the platform among traditional intermediaries such as delivery partners. In the second step, platform governance aimed at **value system expansion**, which involved simulating platform use among various partners and customers. Finally, the **ecosystem expansion** was facilitated by opening up the platform interfaces, promoting interoperability between different platform services as well as creating an open marketplace for new partners to deploy their value-added services.

The industrial digital platform development requires the alignment of strategic priorities within traditional supply chain partners. Consequently, platform sponsors' initial **value chain expansion** started with their delivery partners because they historically played a key role in delivering high value to customers. Accordingly, close collaboration with a network of delivery partners was a necessary step to ensure that the early version of the platform was working flawlessly. Our informants highlighted that the aim was two-fold: first, to allow *subsidized access to the platform*, which signaled partners that the platform offers opportunities for improved collaboration and future gains through platform service delivery; and second, *training delivery partners* that focused on enhancing the digital capabilities of the delivery network and ensuring that partners were equipped for the digital future. Delivery partners were instrumental in delivering tailored platform services to customers. Moreover, delivery partners were trained on how platform services could add value to customers' businesses. A dealer development manager described:

*We have been working for a long time on educating dealers about the potential of our (platform name) system and how to use it in their service processes. Some dealers are more progressive in picking this up and others*

need more support from us. It is still a conservative industry and there can be resistance. But, we realize that we need our dealers on board to deliver increased customer value from our digital investments.

The second phase involved engaging additional partners on the supply side and, more importantly, engaging customers on the demand-side. The **value system expansion** facilitated platform adoption, scaling of platform services, and sharing learnings across different customers' contexts. In order to support *platform adoption among customers*, platform sponsors granted platform services free of charge to all new machines for a limited period of time. Additionally, one executive shared with us that re-evaluating customer segments improved platform adoption. Platform sponsors tried to match platform services with specific customer's needs and technological readiness. For instance, customer segmentation assessment included total machine fleet size, adoption of new ways of working and technologies, analytical capabilities, operational behavior, and the customer's reliance on delivery partners. Simultaneously, platform sponsors supported *platform adoption among partners*. For instance, delivery partners actively participated in the platform service development and the platform service sales process. They bundled different platform services into tailor-made solutions and supported customers on an ongoing basis. On the other hand, the platform enabled delivery partners to monitor customers remotely and take action upon identified deviations. With this approach, problems were diagnosed early and delivery partners' technicians would arrive at the customer's location with the right tools, spare parts, and expertise. This increased the frequency of first-visit fixes, which often happened before the customer had realized that there was a problem. One executive explained:

*During the early years, we interacted with all customers, but we soon realized that we needed to focus on certain customer segments. For example, we had customers that would buy new machines with our digital platform, but turn off most of its features due to misinformation and lower readiness. So, we started focusing on a few progressive customers and we were able to increase their operational productivity. Such success stories helped us to fine-tune our digital offerings and also encouraged other customers to use the platform.*

Also, platform sponsors started *collaborating with selected third parties* in the areas of digital technology development that were outside their core competencies. Certain partners were more important for further advancement of the platform, such as technology partners and specialized service partners. Technology partners were necessary to fast-track certain API developments while their brands added quality assurance to conservative customers. On the other hand, specialized service partners ensured that the touchpoints with customers are well-organized and managed. Consequently, platform sponsors facilitated the onboarding of specialized service partners to ensure quality service to customers. One manager explained:

*We only develop technology if it's absolutely necessary. The goal for us is to utilize what we have, to start with. Then, of course, there will be a need to develop digital solutions that put demands on the hardware, the software and the back-office. In many cases, we can solve this with the partners that we have. At the same point in time, we realized that we will not become the next Microsoft or SAP, so in this area we don't look that much to actually develop our own competence; we should be able to understand what we need and then use third parties and partners to create what we need.*

The third phase involved **ecosystem expansion**. The industrial context brings a high degree of specialization and long-lasting products with high switching costs. Therefore, it was highly unlikely that any platform sponsor would follow the winner-takes-all approach. Instead of focusing on achieving platform dominance, platform sponsors committed to expanding their value creation possibilities for customers through collaboration with partners and industry-level competing platforms. For instance, one informant reported that their customers have a wide product portfolio, namely industrial equipment from different manufacturers, which required *open interfaces* to competitors' products. One manager elaborated:

*We don't want to be naïve and expect that we can convince all our customers to sign up for our platform and digital services. In certain customer segments, such as medium-sized open pits, we can be the dominant machine provider and site optimization solution provider, but in the majority of other cases, we are part of a mixed fleet. In these situations, being considered as the open digital platform provider is a strategic decision.*

Next, *promoting interoperability* between different platform services developed by various partners was important from the customers' point of view. In some cases, interoperability was critical for the configuration of more advanced platform services, such as autonomous services. In fact, the digital landscape created a layer of coopetition; the platform ecosystem allowed competing firms to complement each other's platform services depending on the structure of their customers' industrial assets. One informant elaborated:

*How should we configure solution offers? Because we're looking from the customer's point of view and we need to understand trade-offs. We need to start looking at what we are offering today and what we need to offer in the future. Basically, the question is 'what is our solution portfolio?' and 'what are the competitors' solution offers in those areas?' and 'what are our capabilities, opportunities and data strengths today?'*

*Our internal analysis suggests that those equipment providers that are "playing with open cards", offering an open platform that supports interoperability, will be the winners, unless some company has a significantly superior platform technologically, which is not the case in the industry.*

The ecosystem expansion showed its full generative potential with specialized third-party applications that used the platform's open interfaces to deploy their value-adding services. In several cases, customers merged services from several independent providers with the platform sponsor's services. In other cases, the platform represented an *open marketplace* and an opportunity for technology startups to create additional value for customers with innovative solutions. Thus, the ecosystem expansion has redefined industry boundaries, merging the construction equipment industry with the high-tech industry. One informant set the ambition of their platform ecosystem as high as Apple's ecosystem. He explained:

*We are increasingly thinking like a technological company, such as Apple. They provide the hardware phone but also generate revenue and create customer value through their App Store. We want to create a construction equipment app store for our customers. Such an open marketplace will have our internally developed recommended apps, but also third-party apps. We are committed to this vision of developing an open marketplace, but this is a challenging undertaking.*

*4.4. Expanding the platform value of industrial digital platforms*

In this section, we propose a framework for the industrial digital platform evolution (see Fig. 2). The framework illustrates how platform architecture, platform services, and platform governance co-evolve. Moreover, we find that all three dimensions mirror each other. The framework further elaborates three platform archetypes that we label product platform, supply chain platform, and platform ecosystem, as well as the underlying innovation mechanism for each archetype.

The first phase in the evolution of industrial digital platforms outlined **product platforms**. The product platform was a cornerstone phase for platform sponsors, where they developed the platform core with a machine-centric data, basic data analysis, and ensured value chain partnership to effectively deliver monitoring services. Monitoring services generated a limited platform value through extended **search depth** mechanism within a specific data source. In particular, monitoring service development hinged on *increased data volume* and *detecting errors* within a machine-centric data. According to Global Product Manager:

*With intelligent equipment, we are able to capture "low hanging fruits".*

The next phase delineated **supply chain platforms**. The supply chain platform increased platform architecture functionalities with a fleet data, advanced use of data analysis, and strengthened partnership with both partners and customers that resulted in optimization services. In addition to search depth, optimization services extended the platform value through **search breadth** mechanism. The search breadth mechanism allowed analysis of incomparably *higher data volume and data variety* and *identifying patterns* within larger data sets. Telematics manager illustrated:

> We started to develop the [service name] with the main feature of generating load weighing data, among others. Then we started to realize that we can actually start looking at productivity data. And, we realized that we have the opportunity to correlate different things in a more valuable way than we did before. So, we started looking at uptime, what we could do there. We realized that we have other systems data that we could correlate. We looked at different types of data: we looked at statistical data based on quality claims and warranty data. We created a set of data that could be used. When the data were correlated with these patterns of claims for specific systems and machines, we could generate proactive notifications. It was not that long ago that we launched it, two years ago, and we are now also looking at improving our platform in different ways, to become more advanced. Also, we are looking at the productivity and load weighing data, how we can use it to correlate that with other data.

Finally, the most advanced stage covered **platform ecosystems**. The platform ecosystem additionally increased platform architecture functionalities that included AI-driven data analysis and opened interfaces to diverse partners that collectively enabled autonomous services. Autonomous services expanded the platform value through **recombination** mechanism. The recombination mechanism allowed *utilizing micro-services to configure novel solutions*. Portfolio manager explained the logic of recombination:

> It comes down to what we have to do in the future; if we use micro-services to configure the solutions, we can see this as kind of components of solutions.
> 
> Before, we used to have our machines, our individual services, and if we combine these, we get solutions … but we are moving away from that … we say we have our portfolio of machines, we have our portfolio of individual services (which are mostly related to uptime), but then you have also digital solutions that leverage both of these other portfolios … we sell these digital solutions as an individual entity. It's a modular approach.
> 
> Our solution portfolio for machines are very diverse and developed. This allows us to bundle many smaller services into a customized solution.

## 5. Discussion

The present study aims to propose a framework for the platform evolution in the industrial business-to-business context. In doing so, we offer a holistic perspective that encompasses both technological architectures and platform governance with broader implications for expanding the platform value. The study demarcates three platform archetypes that we label product platform, supply chain platform, and platform ecosystem. Therefore, the framework developed here extends the literature on platform ecosystem and digital servitization in several ways.

First, the present study shows that digitalization does not only facilitates servitization strategies but also enables new platform-driven business models that require reconsidering organizational boundary choices and entail greater dependence on the surrounding digital ecosystem (Huikkola et al., 2020; Kohtamäki et al., 2019). However, due to high specialization in the industrial sector, we argue for the necessity for manufacturers to become platform sponsors and work collaboratively with their partners and customers to develop advanced platform services around their industrial assets (Sklyar et al., 2019; Tronvoll et al., 2020). In particular, the platform ecosystem approach deems essential for the development of higher-order autonomous platform services (Iansiti and Lakhani, 2020; Sandebrg et al., 2020; Svahn et al., 2017).

Second, the study argues that the technological maturity of the platform architecture is associated with specific features of platform services. In order to shift from simple monitoring platform services to more advanced autonomous platform services, platform sponsors need to make gradual investments in the platform architecture to expand platform functionalities (Thomas et al., 2014). We also present how architectural control decisions on boundary resources allow to initially control and later promote complementary extensions to the platform core (Ghazawneh and Henfridsson, 2013). Moreover, we explicate how this process links to specific platform service development. Thus, study contributes to the literature on platform architecture and digital infrastructure (Constantinides et al., 2018; Tilson et al., 2010) with implications for the literature on platform openness and platform generativity (Cenamor and Frishammar, 2021; Cennamo and Santaló, 2019a).

Third, the study shows how the platform architecture development and associated platform governance mechanisms co-evolve. In general, the industrial digital platform evolution usually starts off with an internal focus, and gradually becomes more open and inclusive for other partners. Moreover, the industrial B2B context shows that the platform sponsor would first reach out to its traditional supply chain partners on the supply side (Randhawa et al., 2018) before engaging with customers and other complementors on the demand-side (Sjödin et al., 2020b). Thus, the study further elaborates the antecedents of complementor selection in the evolution of platform ecosystems (Hou and Shi, 2020; McIntyre and Srinivasan, 2017). Moreover, the study explores governance mechanisms in the early phases of the platform ecosystem and unpacks dynamics associated with the platform ecosystem formation (Hannah and Eisenhardt, 2018; Rietveld and Schilling, 2020). Finally, in contrast to the B2C context where the winner-takes-all approach prevails, we argue that the B2B context is vastly different. We display that the high switching cost of the industrial equipment as well as narrow platform scope and platform size would encourage platform coopetition rather than platform competition (Cennamo, 2019; Kretschmer et al., 2020; Pellizzoni et al., 2019).

Fourth, building on the existing platform literature, this study argues that successful industrial digital platform evolution hinges on mirroring platform architecture, platform services, and platform governance (Colfer and Baldwin, 2016). In contrast to the structuration ecosystem views (Adner, 2017; Jacobides et al., 2018), we present cases of the ecosystem co-evolution (Hou and Shi, 2020). We account that the platform sponsor's technological choices, existing supply chain relationships, customer readiness as well as construction equipment market attributes largely affect the platform ecosystem co-evolution. Consequently, we contribute to the co-evolution perspective of platform ecosystem literature.

Finally, the study explicates the underlying innovation mechanisms for three platform archetypes. More specifically, the study presents empirical evidence of how digitalization unlocks innovation opportunities through platform service development (Nambisan et al., 2017; Yoo et al., 2012). Building on digital innovation literature (Lanzolla et al., 2020), we illustrate how digitally-enabled search depth, search breadth, and recombination extend the platform value. Thus, the study contributes to the literature on digital generativity (Kallinikos et al., 2013) and innovation mechanisms behind the platform service discovery (Dattée et al., 2018; Hou and Shi, 2020).

### 5.1. Future research and limitations

This study is not without limitations that call for further research. Though research has begun to emerge on industrial digital platforms, it is still a nascent research field. First, the winner-takes-all mantra from the B2C context is not transferrable to the B2B context. This also implies that the existing platform literature have limited generalizability and transferability to industrial digital platforms. Second, this study acknowledges that scholars should put more focus specifically on the early phases of the evolution of platform ecosystems. Third, industrial digital platforms offer a fruitful context to study how actor-specific data contributes to the development of platform governance and how platform governance mechanisms evolve. Fourth, industrial digital platforms offer a unique context to study the selective promotion of complementors and the transition from closed to open platform ecosystems. Finally, we invite scholars to further explore digital innovation mechanisms that contribute to the platform service discovery.

## 5.2. Managerial implications

In managerial terms, this paper makes several important contributions. First, the paper identifies the critical dimension in the evolution of industrial digital platforms. Prospective platform sponsors are required to simultaneously manage platform architecture, platform services, and platform governance. Second, the study highlights the investments and managerial steps required to transition from monitoring services to autonomous services. Third, we present strategic interactions between different actors in the B2B context and responsibilities of a platform sponsor role. Fourth, access to quality data is a critical step for the development of industrial digital platforms. For instance, platform sponsors often give away telematics systems for free in order to generate quality data for the development of platform services. Finally, platform services need to create a business case for both platform sponsors and customers.

**Acknowledgment**

We gratefully acknowledge the funding support from The Sweden's Innovation Agency (Vinnova) (ref: 2019-04700 and 2018-05295) and a Swedish Research Council for Sustainable Development (Formas) (ref: 2018-01417) to make this research possible.

**References**

Adner, R., 2017. Ecosystem as structure. J. Manag. 43, 39–58. https://doi.org/10.1177/0149206316678451.

Adner, R., Chen, J., Zhu, F., 2019. Frenemies in platform markets: heterogeneous profit foci as drivers of compatibility decisions. Manage. Sci. 66 (6), 2432–2451. https://doi.org/10.1287/mnsc.2019.3327.

Agarwal, R., Tiwana, A., 2015. Editorial—evolvable systems: through the looking glass of IS. Inf. Syst. Res. 26, 473–479. https://doi.org/10.1287/isre.2015.0595.

Alaimo, C., Kallinikos, J., Valderrama, E., 2020. Platforms as service ecosystems: lessons from social media. J. Inf. Technol. 35, 25–48. https://doi.org/10.1177/0268396219881462.

Alvesson, M., 2011. Interpreting Interviews. SAGE Publications Ltd. https://doi.org/10.4135/9781446268353, 1 Oliver's Yard, 55 City Road, London EC1Y 1SP United Kingdom.

Arnold, C., Kiel, D., Voigt, K.-I., 2016. How the Industrial Internet of Things changes business models in different manufacturing industries. Int. J. Innovat. Manag. 20 (08), 1640015. https://doi.org/10.1142/S1363919616400156.

Autio, E., Nambisan, S., Thomas, L.D.W., Wright, M., 2018. Digital affordances, spatial affordances, and the genesis of entrepreneurial ecosystems. Strateg. Entrep. J. 12 (1), 72–95. https://doi.org/10.1002/sej.1266.

Basaure, A., Vesselkov, A., Töyli, J., 2020. Internet of things (IoT) platform competition: consumer switching versus provider multihoming. Technovation 90, 102101. https://doi.org/10.1016/j.technovation.2019.102101.

Bilgeri, D., Gebauer, H., Fleisch, E., Wortmann, F., 2019. Driving process innovation with IoT field data. MIS Q. Exec. 18, 191–207. https://doi.org/10.17705/2msqe.00016.

Björkdahl, J., 2020. Strategies for digitalization in manufacturing firms. Calif. Manag. Rev. 62, 17–36. https://doi.org/10.1177/0008125620920349.

Bozan, K., Lyytinen, K., Rose, G.M., 2021. How to transition incrementally to microservice architecture. Commun. ACM 64 (1), 79–85. https://doi.org/10.1145/3378064.

Braun, V., Clarke, V., 2006. Using thematic analysis in psychology. Qual. Res. Psychol. 3, 77–101. https://doi.org/10.1191/1478088706qp063oa.

Broekhuizen, T.L.J., Emrich, O., Gijsenberg, M.J., Broekhuis, M., Donkers, B., Sloot, L.M., 2021. Digital platform openness: drivers, dimensions and outcomes. J. Bus. Res. 122, 902–914. https://doi.org/10.1016/j.jbusres.2019.07.001.

Cenamor, J., Frishammar, J., 2021. Openness in platform ecosystems: innovation strategies for complementary products. Res. Pol. 50, 104148. https://doi.org/10.1016/j.respol.2020.104148.

Cenamor, J., Rönnberg Sjödin, D., Parida, V., 2017. Adopting a platform approach in servitization: leveraging the value of digitalization. Int. J. Prod. Econ. 192, 54–65. https://doi.org/10.1016/j.ijpe.2016.12.033.

Cennamo, C., 2019. Competing in digital markets: A platform-based perspective. Acad. Manag. Perspect. https://doi.org/10.5465/amp.2016.0048. In press.

Cennamo, C., Dagnino, G.B., Di Minin, A., Lanzolla, G., 2020. Managing digital transformation: scope of transformation and modalities of value Co-generation and delivery. Calif. Manag. Rev. 62, 5–16. https://doi.org/10.1177/0008125620942136.

Cennamo, C., Santaló, J., 2019. Generativity tension and value creation in platform ecosystems. Organ. Sci. 30, 617–641. https://doi.org/10.1287/orsc.2018.1270.

Colfer, L.J., Baldwin, C.Y., 2016. The mirroring hypothesis: theory, evidence, and exceptions. Ind. Corp. Change 25, 709–738. https://doi.org/10.1093/icc/dtw027.

Constantinides, P., Henfridsson, O., Parker, G.G., 2018. Introduction—platforms and infrastructures in the digital age. Inf. Syst. Res. 29, 381–400. https://doi.org/10.1287/isre.2018.0794.

Crainic, T.G., Gendreau, M., Potvin, J.-Y., 2009. Intelligent freight-transportation systems: assessment and the contribution of operations research. Transport. Res. C Emerg. Technol. 17, 541–557. https://doi.org/10.1016/j.trc.2008.07.002.

Dattée, B., Alexy, O., Autio, E., 2018. Maneuvering in Poor Visibility: How Firms Play the Ecosystem Game when Uncertainty is High. Acad. Manag. J. 61 (2), 466–498. https://doi.org/10.5465/amj.2015.0869.

Edmondson, A.C., Mcmanus, S.E., 2007. Methodological fit in management field research. Acad. Manag. Rev. 32, 1155–1179. https://doi.org/10.5465/AMR.2007.26586086.

Eisenhardt, K.M., 1989. Building theories from case study research. Acad. Manag. Rev. 14, 532–550. https://doi.org/10.5465/AMR.1989.4308385.

Eisenhardt, K.M., Graebner, M.E., 2007. Theory building from cases: opportunities and challenges. Acad. Manag. J. 50, 25–32. https://doi.org/10.5465/AMJ.2007.24160888.

Eisenmann, T.R., 2008. Managing proprietary and shared platforms. Calif. Manag. Rev. 50, 31–53. https://doi.org/10.2307/41166455.

Enkel, E., Bogers, M., Chesbrough, H., 2020. Exploring open innovation in the digital age: a maturity model and future research directions. R D Manag. 50, 161–168. https://doi.org/10.1111/radm.12397.

Fontana, A., Frey, J.H., 1998. Interviewing: the art of science. In: Denzin, N.K., Lincoln, Y.S. (Eds.), Collecting and Interpreting Qualitative Materials. Sage Publications Ltd, Thousand Oaks, pp. 47–78.

Gawer, Annabelle, 2020. Digital platforms' boundaries: The interplay of firm scope, platform sides, and digital interfaces. Long Range Plann, 102045.APA.

Gawer, A., 2014. Bridging differing perspectives on technological platforms: toward an integrative framework. Res. Pol. 43, 1239–1249. https://doi.org/10.1016/j.respol.2014.03.006.

Gawer, A., 2009. Platform dynamics and strategies: from products to services. In: Platforms, Markets and Innovation. Edward Elgar Publishing. https://doi.org/10.4337/9781849803311.00009.

Gawer, A., Henderson, R., 2007. Platform owner entry and innovation in complementary markets: evidence from intel. J. Econ. Manag. Strat. 16 (1), 1–34. https://doi.org/10.1111/j.1530-9134.2007.00130.x.

Gebauer, H., Paiola, M., Saccani, N., Rapaccini, M., 2020. Digital servitization: crossing the perspectives of digitization and servitization. Ind. Market. Manag. https://doi.org/10.1016/j.indmarman.2020.05.011. In press.

Ghazawneh, A., Henfridsson, O., 2013. Balancing platform control and external contribution in third-party development: the boundary resources model. Inf. Syst. J. 23, 173–192. https://doi.org/10.1111/j.1365-2575.2012.00406.x.

Ghezzi, A., Cavallo, A., 2020. Agile Business Model Innovation in Digital Entrepreneurship: Lean Startup Approaches. J. Business Res. 110 (March), 519–537. https://doi.org/10.1016/j.jbusres.2018.06.013.

Gioia, D.A., Corley, K.G., Hamilton, A.L., 2013. Seeking qualitative rigor in inductive research: notes on the Gioia methodology. Organ. Res. Methods 16, 15–31. https://doi.org/10.1177/1094428112452151.

Glaser, B.G., Strauss, A., 1967. The Discovery of Grounded Theory: Strategies for Qualitative Research. Aldine Publishing Co., Chicago, IL.

Granstrand, O., Holgersson, M., 2020. Innovation ecosystems: a conceptual review and a new definition. Technovation 90–91, 102098. https://doi.org/10.1016/j.technovation.2019.102098.

Gregory, R.W., Henfridsson, O., Kaganer, E., Kyriakou, H., 2020. The role of artificial intelligence and data network effects for creating user value. Acad. Manag. Rev. amr.2019.0178. https://doi.org/10.5465/amr.2019.0178. In press.

Gulati, R., Puranam, P., Tushman, M., 2012. Meta-organization design: rethinking design in interorganizational and community contexts. Strat. Manag. J. 33, 571–586. https://doi.org/10.1002/smj.1975.

Haefner, N., Wincent, J., Parida, V., Gassmann, O., 2021. Artificial intelligence and innovation management: a review, framework, and research agenda☆. Technol. Forecast. Soc. Change 162, 120392. https://doi.org/10.1016/j.techfore.2020.120392.

Hanelt, A., Bohnsack, R., Marz, D., Antunes, C., 2020. A systematic review of the literature on digital transformation: insights and implications for strategy and organizational change. J. Manag. Stud. https://doi.org/10.1111/joms.12639. In press.

Hannah, D.P., Eisenhardt, K.M., 2018. How firms navigate cooperation and competition in nascent ecosystems. Strateg. Manag. J. 39 (12), 3163–3192. https://doi.org/10.1002/smj.2750.

Henfridsson, O., Bygstad, B., 2013. The generative mechanisms of digital infrastructure evolution. MIS Q. 37, 907–931. https://doi.org/10.25300/MISQ/2013/37.3.11.

Hilbolling, S., Berends, H., Deken, F., Tuertscher, P., 2020. Complementors as connectors: managing open innovation around digital product platforms. R D Manag. 50, 18–30. https://doi.org/10.1111/radm.12371.

Hou, H., Shi, Y., 2020. Ecosystem-as-structure and ecosystem-as-coevolution: a constructive examination. Technovation 102193. https://doi.org/10.1016/j.technovation.2020.102193. In press.

Huikkola, T., Rabetino, R., Kohtamäki, M., Gebauer, H., 2020. Firm boundaries in servitization: interplay and repositioning practices. Ind. Market. Manag. 90, 90–105. https://doi.org/10.1016/j.indmarman.2020.06.014.

Iansiti, M., Lakhani, K., 2020. Competing in the age of AI. Harvard Buiness Rev 98, 3–9.

Jacobides, M.G., Cennamo, C., Gawer, A., 2018. Towards a theory of ecosystems. Strat. Manag. J. 39, 2255–2276. https://doi.org/10.1002/smj.2904.

Jocevski, M., 2020. Blurring the lines between physical and digital spaces: business model innovation in retailing. Calif. Manag. Rev. 63, 99–117. https://doi.org/10.1177/0008125620953639.

Kallinikos, J., Aaltonen, A., Marton, A., 2013. The ambivalent ontology of digital artifacts. MIS Q. Manag. Inf. Syst. 357–370. https://doi.org/10.25300/MISQ/2013/37.2.02.

Kiel, D., Arnold, C., Voigt, K.I., 2017. The influence of the Industrial Internet of Things on business models of established manufacturing companies – a business level perspective. Technovation 68, 4–19. https://doi.org/10.1016/j.technovation.2017.09.003.

Kieu, T., Yang, B., Jensen, C.S., 2018. Outlier detection for multidimensional time series using deep neural networks. In: Proceedings - IEEE International Conference on Mobile Data Management. https://doi.org/10.1109/MDM.2018.00029.

Kohtamäki, M., Parida, V., Oghazi, P., Gebauer, H., Baines, T., 2019. Digital servitization business models in ecosystems: a theory of the firm. J. Bus. Res. 104, 380–392. https://doi.org/10.1016/j.jbusres.2019.06.027.

Kohtamäki, M., Parida, V., Patel, P.C., Gebauer, H., 2020. The relationship between digitalization and servitization: the role of servitization in capturing the financial potential of digitalization. Technol. Forecast. Soc. Change 151, 119804. https://doi.org/10.1016/j.techfore.2019.119804.

Koutsikouri, D., Lindgren, R., Henfridsson, O., Rudmark, D., 2018. Extending digital infrastructures: a typology of growth tactics. J. Assoc. Inf. Syst. Online 1001–1019. https://doi.org/10.17705/1jais.00517.

Kretschmer, T., Leiponen, A., Schilling, M., Vasudeva, G., 2020. Platform ecosystems as metaorganizations: implications for platform strategies. Strateg. Manag. J. https://doi.org/10.1002/smj.3250. In press.

Kvale, S., 1996. InterViews:An Introduction to Qualitative Research Interviewing. Sage Publications Ltd, London.

Lanzolla, G., Pesce, D., Tucci, C.L., 2020. The digital transformation of search and recombination in the innovation function: tensions and an integrative framework*. J. Prod. Innov. Manag. https://doi.org/10.1111/jpim.12546. In press.

Leminen, S., Rajahonka, M., Wendelin, R., Westerlund, M., 2020. Industrial internet of things business models in the machine-to-machine context. Ind. Market. Manag. 84, 298–311. https://doi.org/10.1016/j.indmarman.2019.08.008.

Leminen, S., Rajahonka, M., Westerlund, M., Wendelin, R., 2018. The future of the Internet of Things: toward heterarchical ecosystems and service business models. J. Bus. Ind. Market. 33, 749–767. https://doi.org/10.1108/JBIM-10-2015-0206.

Lenka, S., Parida, V., Wincent, J., 2017. Digitalization capabilities as enablers of value Co-creation in servitizing firms. Psychol. Market. 34, 92–100. https://doi.org/10.1002/mar.20975.

McAfee, A., Brynjolfsson, E., 2012. Big Data. The management revolution. Harvard Buiness Rev 90, 61–68. https://doi.org/10.1007/s12599-013-0249-5.

Mcintyre, D., 2019. Beyond a 'winner-takes-all' strategy for platforms. MIT Sloan Manag. Rev.

McIntyre, David P., Srinivasan, Arati, 2017. Networks, platforms, and strategy: Emerging views and next steps. Strat. Manage. J. 38 (1), 141–160.

Miles, M., Huberman, A., 1994. Qualitative data analysis: an expanded sourcebook. In: Qualitative Data Analysis: an Expanded Sourcebook, second ed. SAGE Publications, Inc.

Miller, C.C., Cardinal, L.B., Glick, W.H., 1997. Retrospective reports in organizational research: a reexamination of recent evidence. Acad. Manag. J. 40, 189–204. https://doi.org/10.2307/257026.

Miller, D.C., Salkind, N.J., 2002. Handbook of Research Design and Social Measurement, sixth ed. SAGE Publications, Inc, Thousand Oaks, CA.

Nambisan, S., Lyytinen, K., Majchrzak, A., Song, M., 2017. Digital innovation management: reinventing innovation management research in a digital world. MIS Q. 41 (1) https://doi.org/10.25300/MISQ/2017/41:1.03.

Nambisan, S., Wright, M., Feldman, M., 2019. The digital transformation of innovation and entrepreneurship: progress, challenges and key themes. Res. Pol. 48, 103773. https://doi.org/10.1016/j.respol.2019.03.018.

Ng, I.C.L., Wakenshaw, S.Y.L., 2017. The Internet-of-Things: review and research directions. Int. J. Res. Market. 34, 3–21. https://doi.org/10.1016/j.ijresmar.2016.11.003.

Paiola, M., Gebauer, H., 2020. Internet of things technologies, digital servitization and business model innovation in BtoB manufacturing firms. Ind. Market. Manag. 89, 245–264. https://doi.org/10.1016/j.indmarman.2020.03.009.

Panico, C., Cennamo, C., 2020. User preferences and strategic interactions in platform ecosystems. Strat. Manag. J. https://doi.org/10.1002/smj.3149.

Parker, G., Van Alstyne, M., Choudary, S.P., 2016. Platform Revolution. W.W. Norton & Company.

Paschou, T., Rapaccini, M., Adrodegari, F., Saccani, N., 2020. Digital servitization in manufacturing: a systematic literature review and research agenda. Ind. Market. Manag. 89, 278–292. https://doi.org/10.1016/j.indmarman.2020.02.012.

Pellizzoni, E., Trabucchi, D., Buganza, T., 2019. Platform strategies: how the position in the network drives success. Technol. Anal. Strat. Manag. 31 (5), 579–592. https://doi.org/10.1080/09537325.2018.1524865.

Porter, M.J.E., Heppelmann, J.E., 2014. How smart, connected products are transforming companies. Harv. Bus. Rev. 92, 64. https://doi.org/10.1007/s13398-014-0173-7.2.

Raff, S., Wentzel, D., Obwegeser, N., 2020. Smart products: conceptual review, synthesis, and research directions. J. Prod. Innovat. Manag. 37, 379–404. https://doi.org/10.1111/jpim.12544.

Rajala, R., Brax, S.A., Virtanen, A., Salonen, A., 2019. The next phase in servitization: transforming integrated solutions into modular solutions. Int. J. Oper. Prod. Manag. 39 (5), 630–657. https://doi.org/10.1108/ijopm-04-2018-0195.

Randhawa, K., Wilden, R., Gudergan, S., 2018. Open service innovation: the role of intermediary capabilities. J. Prod. Innovat. Manag. 35, 808–838. https://doi.org/10.1111/jpim.12460.

Rietveld, J., Schilling, M.A., 2020. Platform competition: a systematic and interdisciplinary review of the literature. J. Manag. https://doi.org/10.2139/ssrn.3706452. In press.

Rietveld, J., Schilling, M.A., Bellavitis, C., 2019. Platform strategy: managing ecosystem value through selective promotion of complements. Organ. Sci. 30, 1232–1251. https://doi.org/10.1287/orsc.2019.1290.

Rifkin, J., 2015. The Zero Marginal Cost Society: the Internet of Things, the Collaborative Commons, and the Eclipse of Capitalism. Griffin.

Rowley, J., 2002. Using case studies in research. Manag. Res. News 25, 16–27. https://doi.org/10.1108/01409170210782990.

Saadatmand, F., Lindgren, R., Schultze, U., 2019. Configurations of platform organizations: Implications for complementor engagement. Research Policy 48 (8). https://doi.org/10.1016/j.respol.2019.03.015.

Sandebrg, J., Holmstrom, J., Lyytinen, K., 2020. Digitization and phase transitions in platform organizing logics: evidence from the process automation industry. MIS Q. 44, 129–153. https://doi.org/10.25300/MISQ/2020/14520.

Savino, T., Messeni Petruzzelli, A., Albino, V., 2017. Search and recombination process to innovate: a review of the empirical evidence and a research agenda. Int. J. Manag. Rev. 19, 54–75. https://doi.org/10.1111/ijmr.12081.

Schroeder, A., Naik, P., Ziaee Bigdeli, A., Baines, T., 2020. Digitally enabled advanced services: a socio-technical perspective on the role of the internet of things (IoT). Int. J. Oper. Prod. Manag. 40 (7/8), 1243–1268. https://doi.org/10.1108/IJOPM-03-2020-0131.

Sestino, A., Prete, M.I., Piper, L., Guido, G., 2020. Internet of Things and Big Data as enablers for business digitalization strategies. Technovation 98, 102173. https://doi.org/10.1016/j.technovation.2020.102173.

Sjödin, D., Parida, V., Jovanovic, M., Visnjic, I., 2020a. Value creation and value capture alignment in business model innovation: a process view on outcome-based business models. J. Prod. Innovat. Manag. 37, 158–183. https://doi.org/10.1111/jpim.12516.

Sjödin, D., Parida, V., Kohtamäki, M., Wincent, J., 2020b. An agile co-creation process for digital servitization: a micro-service innovation approach. J. Bus. Res. 112, 478–491.

Sklyar, A., Kowalkowski, C., Tronvoll, B., Sörhammar, D., 2019. Organizing for digital servitization: a service ecosystem perspective. J. Bus. Res. 104, 450–460. https://doi.org/10.1016/j.jbusres.2019.02.012.

Strauss, A., Corbin, J., 2015. Basics of Qualitative Research, Handbook of Qualitative Research. Sage, Thousand Oaks, CA.

Suppatvech, C., Godsell, J., Day, S., 2019. The roles of internet of things technology in enabling servitized business models: a systematic literature review. Ind. Market. Manag. 82, 70–86. https://doi.org/10.1016/j.indmarman.2019.02.016.

Svahn, F., Mathiassen, L., Lindgren, R., 2017. Embracing digital innovation in incumbent firms: how Volvo cars managed competing concerns. MIS Q. 41, 239–253. https://doi.org/10.25300/MISQ/2017/41.1.12.

Thomas, L.D.W., Autio, E., Gann, D.M., 2014. Architectural leverage: putting platforms in context. Acad. Manag. Perspect. 28, 198–219. https://doi.org/10.5465/amp.2011.0105.

Tilson, D., Lyytinen, K., Sørensen, C., 2010. Digital infrastructures: the missing IS research agenda. Inf. Syst. Res. 21 (4), 748–759. https://doi.org/10.1287/isre.1100.0318.

Tiwana, A., 2014. Platform ecosystems: aligning architecture, governance, and strategy, platform ecosystems: aligning architecture, governance, and strategy. https://doi.org/10.1016/C2012-0-06625-2.

Tiwana, A., Konsynski, B., Bush, A.A., 2010. Research commentary —platform evolution: coevolution of platform architecture, governance, and environmental dynamics. Inf. Syst. Res. 21, 675–687. https://doi.org/10.1287/isre.1100.0323.

Tronvoll, B., Sklyar, A., Sörhammar, D., Kowalkowski, C., 2020. Transformational shifts through digital servitization. Ind. Market. Manag. 89, 293–305. https://doi.org/10.1016/j.indmarman.2020.02.005.

Vendrell-Herrero, F., Bustinza, O.F., Parry, G., Georgantzis, N., 2017. Servitization, digitization and supply chain interdependency. Ind. Market. Manag. 60, 69–81. https://doi.org/10.1016/j.indmarman.2016.06.013.

Visnjic, I., Jovanovic, M., Neely, A., Engwell, M., 2017. What brings the value to outcome-based contract providers? Value drivers in outcome business models. Int. J. Prod. Econ. 192 (October), 169–181. https://doi.org/10.1016/j.ijpe.2016.12.008.

Visnjic, I., Neely, A., Jovanovic, M., 2018. The path to outcome delivery: Interplay of service market strategy and open business models. Technovation 72-73 (April–May), 46–59. https://doi.org/10.1016/j.technovation.2018.02.003.

Warner, K.S.R., Wäger, M., 2019. Building dynamic capabilities for digital transformation: an ongoing process of strategic renewal. Long. Range Plan. https://doi.org/10.1016/j.lrp.2018.12.001.

Wei, R., Geiger, S., Vize, R., 2019. A platform approach in solution business: how platform openness can be used to control solution networks. Ind. Market. Manag. https://doi.org/10.1016/j.indmarman.2019.04.010.

Weitzman, M.L., 1998. Recombinant growth. Q. J. Econ. 133 (2), 331–360. https://doi.org/10.1162/003355398555595.

Yin, R.K., 2017. Case Study Research: Design and Methods, sixth ed. SAGE Publications, Inc, Thousand Oaks, CA.

Yoo, Y., Boland, R.J., Lyytinen, K., Majchrzak, A., 2012. Organizing for innovation in the digitized world. Organ. Sci. https://doi.org/10.1287/orsc.1120.0771.

Yoo, Y., Henfridsson, O., Lyytinen, K., 2010. The new organizing logic of digital innovation: an agenda for information systems research. Inf. Syst. Res. https://doi.org/10.1287/isre.1100.0322.

Zheng, T., Ardolino, M., Bacchetti, A., Perona, M., 2020. The applications of Industry 4.0 technologies in manufacturing context: a systematic literature review. Int. J. Prod. Res. 1–33. https://doi.org/10.1080/00207543.2020.1824085.

Zhu, F., Furr, N., 2016. Products to platforms: making the leap. Harv. Bus. Rev. April.

Dr. Marin Jovanovic is an assistant professor at the department of operations management at Copenhagen Business School and a visiting scholar at Luleå University of Technology. He received a Ph.D. degree in industrial economics and management from the KTH Royal Institute of Technology and a Ph.D. degree (cum laude) in industrial management from the Universidad Politécnica de Madrid. His research has been published in academic journals, such as Journal of Product Innovation Management, Technovation, International Journal of Production Economics, Journal of Business Research, and Research-Technology Management. His main research revolves around the servitization and digitalization of manufacturing and platform ecosystems in B2B context. Marin has held positions at the ESADE Business School and University of Cambridge.

Prof. David Sjödin is an associate professor of entrepreneurship and innovation at Luleå University of Technology, Sweden and a professor of entrepreneurship and innovation at University of South Eastern Norway. He conducts research on the topics of servitization, digitalization, open innovation, and business model innovation in collaboration with leading global firms and regularly consults industry. He has published 25+ papers in distinguished international journals, including California Management Review, Long Range Planning, Journal of Product Innovation Management, Journal of Business Research, and others. He is the recipient of multiple awards for his research, including the Entrepreneurship Forum Young Researcher Award 2018 for his research on the servitization of industry.

Prof. Vinit Parida is a chaired professor of entrepreneurship and innovation at Luleå University of Technology, Sweden and a professor of entrepreneurship and innovation at University of South Eastern Norway. He is an associate editor for Journal of Business Research in Business-to-Business (B2B) track. He conducts research on the topics of business model innovation, digitalization, circular economy, and organizational capabilities. He has published 80+ papers in distinguished international journals, including Strategic Management Journal, Journal of Management Studies, Industrial Marketing Management, Production and Operation Management, Strategic Entrepreneurship Journal, and others. He is the recipient of multiple awards for his research work.